\documentclass[fleqn,usenatbib]{mnras}

\usepackage[usenames,dvipsnames]{xcolor}

\hypersetup{
    colorlinks = true,
    citecolor = {MidnightBlue},
    linkcolor = {BrickRed},
    urlcolor = {BrickRed},
}

\usepackage{newtxtext,newtxmath}

\usepackage[T1]{fontenc}
\usepackage{ae,aecompl}
\usepackage{multirow}
\allowdisplaybreaks[1]

\usepackage{graphicx}	
\usepackage{bm}
\usepackage{balance}
\usepackage{dcolumn}
\usepackage{amsmath,amsfonts,amssymb,slashed,upgreek,color}
\usepackage{multirow}
\usepackage[mathscr]{eucal}
\usepackage{enumitem}

\usepackage{geometry}
\newcommand\ee{\end{equation}}
\newcommand\be{\begin{equation}}
\newcommand\eea{\end{eqnarray}}
\newcommand\bea{\begin{eqnarray}}
\newcommand{\bV}{\mathbf{V}}

\newcommand{\bn}{\mathbf{n}}

\newcommand{\bk}{\mathbf{k}}

\newcommand{\ks}{k_*}
\newcommand{\nk}{m}
\newcommand{\hxiq}{\hat\xi_2^{\Delta}}
\newcommand{\hxih}{\hat\xi_4^{\Delta}}
\newcommand{\hxid}{\hat\xi_1^{\kappa}}
\newcommand{\xiq}{\xi_2^{\Delta}}
\newcommand{\xih}{\xi_4^{\Delta}}
\newcommand{\xid}{\xi_1^{\kappa}}
\newcommand{\HH}{\mathcal{H}}
\newcommand{\FF}{\mathcal{F}}
\newcommand{\NN}{\mathcal{N}_f}
\newcommand{\hNN}{\hat{\mathcal{N}}_f}

\newcolumntype{C}[1]{>{\centering\arraybackslash}p{#1}}

\title[Null test of the growth rate]{A null test to probe the scale-dependence of the growth of structure as a test of General Relativity}

\author[F. O. Franco et al.]{Felipe Oliveira Franco$^{1}$,
Camille Bonvin$^{1}$, 
Chris Clarkson$^{2,3,4}$
\\\\
$^{1}$D\'epartement de Physique Th\'eorique and Center for Astroparticle Physics, University of Geneva,
CH-1211 Geneva, Switzerland \\
$^{2}$School of Physics \& Astronomy, Queen Mary University of London, London E1 4NS, UK\\
$^{3}$Department of Physics \& Astronomy, University of the Western Cape, Cape Town 7535, South Africa \\
$^{4}$Department of Mathematics \& Applied Mathematics, University of Cape Town, Cape Town 7701, South Africa
}

\date{Accepted XXX. Received YYY; in original form ZZZ}

\pubyear{2019}

\hypersetup{draft}

\begin{document}
\label{firstpage}
\pagerange{\pageref{firstpage}--\pageref{lastpage}}
\maketitle

\begin{abstract}
The main science driver for the coming generation of cosmological surveys is understanding dark energy which relies on testing General Relativity on the largest scales. Once we move beyond the simplest explanation for dark energy of a cosmological constant, the space of possible theories becomes both vast and extremely hard to compute realistic observables. A key discriminator of a cosmological constant, however, is that the growth of structure is scale-invariant on large scales. By carefully weighting observables derived from distributions of galaxies and a dipole pattern in their apparent sizes, we construct a null test which vanishes for any model of gravity or dark energy where the growth of structure is scale-independent. It relies only on very few assumptions about cosmology, and does not require any modelling of the growth of structure. We show that with a survey like DESI a scale-dependence of the order of 10-20 percent can be detected at 3 sigma with the null test, which will drop by a factor of 2 for a survey like the Square Kilometre Array. We also show that the null test is very insensitive to typical uncertainties in other cosmological parameters including massive neutrinos and scale-dependent bias, making this a key null test for dark energy.
\end{abstract}



\section{Introduction}

The large-scale structure (LSS) of the Universe is highly sensitive to the theory of gravity and provides therefore a powerful way of testing for deviations from General Relativity (GR). The standard way to test for modifications of gravity is to measure LSS observables, and confront these measurements with a theoretical modelling which accounts for deviations from GR. This can be done in two complementary ways: the first one consists in calculating observables in a specific model of modified gravity or dark energy, which usually depends on some free parameters, and use observations to place constraints on these parameters. This approach can be used to test specific models, like for example $f(R)$ gravity~\citep{1970MNRAS.150....1B,1980PhLB...91...99S}. The second approach parameterizes deviations from GR directly at the level of the observables. One well-known example is the $\gamma$ parameterization of the growth rate~\citep{Wang_1998,PhysRevD.72.043529}: $f(z)=\Omega_m(z)^\gamma$, where $\gamma$ is a free parameter which takes the value $\gamma\simeq0.55$ in GR and can be directly constrained with LSS observables. In the last decade, various frameworks have been developed, like the Effective Theory of Dark Energy~\citep{Gubitosi:2012hu} and the Parameterized Post-Friedmann approach~\citep{Baker:2012zs}, to combine these two approaches. The goal of these frameworks is to propose parameterizations of deviations from GR that can describe large classes of theories, and whose parameters directly affect LSS observables. These parameterizations provide therefore a consistent way of testing deviations from GR. They suffer however from two limitations. First, to be as general as possible, these parameterizations contain various free functions of time, that cannot all be constrained by observations, and that can therefore not be reconstructed without additional assumptions. Second, even if these parameterizations are very general, they do not account for all possible deviations from $\Lambda$CDM. Hence, by using them, we automatically restrict ourselves to some specific classes of theories. 

In this context, it is important to take a complementary approach, by constructing tests that do not rely on any modelling of the theory of gravity, but that can be used to test one specific property, e.g., the $E_g$ statistics~\citep{Zhang:2007nk,Ghosh:2018ijm}. In this letter, we propose \emph{a null test to probe the scale-independence of the growth of structure} in the linear regime. In $\Lambda$CDM, matter density perturbations grow at the same rate inside the horizon. As a consequence, perturbations at different redshifts are related by a scale-independent function: $\delta(z, \bk)=D_1(z)/D_1(z')\delta(z', \bk)$, with $D_1$ the linear growth~\citep{Dodelson:2003ft}. The continuity equation implies then that the peculiar velocity is related to the density by the growth rate $f(z)=d\ln D_1/d\ln a$, with $a$ the scale factor.
The aim of this letter is to combine LSS observables to construct a null test, $\NN$, which exactly vanishes if and only if $D_1$ and $f$ are scale-independent. We will see that this null test does not require any modelling of deviations from scale-independence. As such it allows us to probe in a model-independent way if structures grow at the same rate at all scales, or if some scales are enhanced or suppressed. Modified theories of gravity generically produce a growth rate which depends on scale~\citep{DeFelice:2011hq}. However, this scale-dependence does not affect modes that are well inside the sound horizon of dark energy, in the regime where the extreme quasi-static approximation is valid~\citep{Gleyzes:2015rua,Sawicki:2015zya}. A detection of  $\NN\neq 0$ would therefore rule out not only $\Lambda$CDM but also all dark energy and modified gravity theories with a growth that differs from $\Lambda$CDM but is scale-independent. Alternatively, a vanishing $\NN$ would put stringent constraints on scale-dependent theories.

To construct our null test we use LSS observables that are sensitive to the growth rate $f$, and we combine them in such a way that the result vanishes if $f$ is scale-independent. The growth rate is related to the galaxy peculiar velocities, which are traditionally measured from redshift-space distortions (RSD)~\citep{Kaiser:1987qv,Hamilton:1997zq}, namely from the monopole, quadrupole and hexadecapole of galaxy clustering. Among these quantities the monopole is the only one which is sensitive to density-density correlations, so we cannot construct a null test by using only these observables. However, an alternative way to measure peculiar velocities has been proposed recently, by looking at their impact on the size of galaxies, i.e.\ by measuring the cosmic convergence~\citep{Bonvin:2008ni,Bolejko:2012uj,Bacon:2014uja}. In particular, \cite{Bonvin:2016dze} showed that peculiar velocities generate a dipolar modulation in the number count-convergence correlation. This effect, called Doppler magnification, has not been measured yet, but its signal-to-noise with a survey like DESI~\citep{Aghamousa:2016zmz}, is expected to reach 37~\citep{Bonvin:2016dze}. Since this effect is sensitive to both the density-velocity correlations and the velocity-velocity correlations, we can combine it with the quadrupole and hexadecapole of RSD to construct our null test.

\section{Methodology}

Redshift surveys map the distribution of galaxies in redshift-space, providing a measurement of the overdensity of galaxies $\Delta(z,\bn)$ at redshift $z$ and in direction $\bn$. The two main contributions to $\Delta$ are given by the matter density fluctuations and RSD.  
In addition, lensing surveys measure the size and luminosity of galaxies, from which one can construct an estimator for the convergence~\citep{Schmidt:2011qj,Casaponsa:2012tq}. The two main contributions are given by~\citep{Bonvin:2008ni}
\be
\kappa(z, \bn)=\int_0^r dr'\frac{r-r'}{2rr'}\Delta_\Omega(\Phi+\Psi)+\left(\frac{1}{r\HH}-1 \right)\bV\cdot\bn\, ,
\ee
where $\Phi$ and $\Psi$ are the metric potentials, $r$ is the radial conformal distance, $\HH$ is the Hubble parameter in conformal time and $\Delta_\Omega$ is the angular Laplacian. The first term is the standard gravitational lensing, whereas the second term is the so-called Doppler magnification. This contribution is due to the fact that a galaxy with a peculiar velocity directed e.g.\ towards the observer, will be further away in real space than a galaxy with no peculiar velocity observed at the same redshift. As a consequence, the first galaxy will appear demagnified with respect to the second one, simply due to its larger distance. 
Note that in both $\Delta$ and $\kappa$ we neglect relativistic effects and magnification bias \citep{Bonvin:2008ni,Yoo:2009au,Bonvin:2011bg,Challinor:2011bk,Jeong:2011as} as these are subdominant in the regime we are interested in.

To construct the null test, we combine three different observables: the quadrupole of $\langle\Delta\Delta\rangle$, $\hat\xi_2^{\Delta}(d, z)$, the hexadecapole of $\langle\Delta\Delta\rangle$, $\hxih(d, z)$, and the dipole of $\langle\Delta\kappa\rangle$, $\hxid(d, z)$. Here, $d$ is the separation between galaxies, and $z$ is the mean redshift of the bin in which the multipoles are measured. 
The lensing contribution in $\kappa$ is negligible in the dipole for $z\leq 0.5$~\citep{Bonvin:2016dze}. The quadrupole, hexadecapole and dipole are therefore all given by combinations of density-velocity correlations and velocity-velocity correlations. In all generality, the evolution of density perturbations can be encoded in a scale-dependent growth function $D_1(z, k)$ such that 
$
\delta(z,\bk)=D_1(z, k)/D_1(z', k)\delta(z',\bk)\, .
$
Due to statistical isotropy, $D_1$ cannot depend on the direction of $\bk$. Using the continuity equation, which is valid in any theory of gravity as long as there is no flow of energy from matter to another component, we obtain for the velocity potential at sub-horizon scale
$
V(z, \bk)=-\HH(z) f(z, k)\delta(z, \bk)/k\, ,
$
where the growth rate $f$ is defined as
$
f(z, k)=\frac{d\ln D_1(z, k)}{d\ln a}\,.
$

In the flat-sky approximation, the mean of 
the quadrupole, hexadecapole and dipole can be written as
\begin{align}
\xiq(d, z)=&-\frac{1}{2\pi^2}\int dk k^2 \left(\frac{4}{3}b(z)f(z, k)+\frac{4}{7}f^2(z, k) \right)\nonumber\\
&\hspace{1.3cm}\times P(k, z)j_2(k d)\, ,\nonumber\\
\xih(d, z)=&\frac{1}{2\pi^2}\int dk k^2 \frac{8}{35}f^2(z, k) P(k, z)j_4(k d)\, ,\nonumber\\
\xid(d, z)=&g(z)\frac{1}{2\pi^2}\int dk k\HH_0 \left(b(z)f(z, k)+\frac{3}{5}f^2(z, k) \right)\nonumber\\
&\hspace{1.3cm}\times P(k, z)j_1(k d)\, ,\label{hatxi}
\end{align}
with $j_\ell$ the spherical Bessel functions, $b(z)$ the bias, $P(k, z)$ the matter power spectrum,
and 
\be
g(z)=\frac{\HH(z)}{\HH_0}\left(1-\frac{1}{r(z)\HH(z)} \right)\,. \label{gz}
\ee

We now construct our null test as
\be
\hNN(d, z)\equiv\frac{\hxiq(d,z)}{\bar \mu_2(d, z)} - \frac{\hxih(d,z)}{\bar \mu_4(d, z)}+\frac{4}{3\bar{g}(z)}\frac{\hxid(d,z)}{\bar \nu_1(d, z)}\, ,\quad\mbox{where} \label{N0def}
\ee
\begin{align}
\mu_\ell(d, z)&=\frac{1}{2\pi^2}\int dk k^2 P(k, z)j_\ell(k d)\, ,\label{mu}\\
\nu_1(d, z)&=\frac{1}{2\pi^2}\int dk k\HH_0 P(k, z)j_1(k d)\, ,\label{nu}
\end{align}
and we denote by a bar all quantities calculated in $\Lambda$CDM. Our null test is therefore a combination of observables $\hxiq, \hxih$ and $\hxid$ that are directly measured from the data, weighted by appropriate coefficients calculated in $\Lambda$CDM. Note that to construct $\hNN$ we do not need \emph{any} modelling of the growth of structure $D_1$ and $f$.

Under which conditions does the mean of $\hNN$ vanish? If the growth rate is scale-independent, then $f(z)$ in Eq.~\eqref{hatxi}  can be taken out of the integrals. Moreover in this case the power spectrum can be related to the one in $\Lambda$CDM by 
\be
\label{eq:Pevol}
P(k, z)=\left(\frac{D_1(z)}{\bar D_1(z)}\right)^2\bar P(k, z)\, ,
\ee
provided that the $\Lambda$CDM model has the same cosmological parameters as the actual Universe: $\Omega_b$, $\Omega_m$, $n_s$ and $h$. The statistical average of $\hNN$, $\NN(d, z)=\langle\hNN(d, z)\rangle$, becomes then
\begin{align}
\NN(d, z)
&=4f(z)\left(\frac{D_1(z)}{\bar D_1(z)}\right)^2\left(\frac{b(z)}{3}+\frac{f(z)}{5} \right)
\left(\frac{g(z)}{\bar g(z)}-1 \right)\, .\label{N0vanish}
\end{align}
When $f$ is scale-independent, there are therefore two additional conditions for $\NN$ to vanish. First $g(z)$ must be the same as the one calculated in $\Lambda$CDM. 
From Eq.~\eqref{gz} we see that $g(z)$ depends only on the evolution of the background, which is currently constrained to follow $\Lambda$CDM up to a very good precision. We will show below that varying $g(z)$ within the 2$\sigma$ region allowed by Planck~\citep{Aghanim:2018eyx} generates a non-zero $\NN$. This value is however negligible compared to the variance of the estimator, that we calculate below. This means that a detection of $\NN\neq0$ cannot be due to our choice of $\bar g(z)$ in Eq.~\eqref{N0def}, except if Planck constraints on the background evolution are incorrect at more than 2$\sigma$. 

The second condition for $\NN$ to vanish is that Eq.~\eqref{eq:Pevol} holds, i.e.\ that 
the cosmological parameters used to calculate $\bar P(k,z)$, namely $\Omega_b$, $\Omega_m$, $n_s$ and $h$~\footnote{Note that the uncertainty in the amplitude of primordial fluctuations, $A_s$, has no impact on $\NN$, since it only rescales the power spectrum by a scale-independent factor.}, are the correct ones.
We  show below that varying these parameters by 2$\sigma$ around the fiducial Planck cosmology modifies Eq.~\eqref{eq:Pevol} and generates consequently a non-zero $\NN$. However the value of $\NN$ in this case is again much smaller than the variance of the estimator. A detection of $\NN\neq0$ can therefore not be due to our choice of cosmological parameters to calculate $\bar \mu_2, \bar\mu_4$ and $\bar\nu_1$ in Eq.~\eqref{N0def}, except if Planck constraints on the parameters $\Omega_b$, $\Omega_m$, $n_s$ and $h$ are incorrect at more than 2$\sigma$. 

To summarise, the null test vanishes whenever the following hold:
\begin{enumerate}[label=(\arabic*),leftmargin=0.6cm,itemsep=0.05cm, topsep=0.08cm]
\item The growth rate of structure $f$ is scale-independent.
\item The background evolution is close to $\Lambda$CDM at redshift $z$, within Planck constraints.
\item The cosmological parameters $\Omega_b$, $\Omega_m$, $n_s$ and $h$ are consistent with Planck constraints.  
\end{enumerate}
From Eq.~\eqref{N0vanish}, we see that under these conditions $\NN$ effectively vanishes, for {\it any form} of the functions $D_1$ and $f$. For example, all Horndeski theories that are consistent with Planck constraints (i.e.\ that have a $\Lambda$CDM-like background) and for which the quasi-static approximation is valid~\citep{Gleyzes:2015rua} have $\NN=0$, even if the growth of structure in these theories differs from $\Lambda$CDM. The fact that in these theories $D_1$ differs from $\bar D_1$ (used to calculate the weights in Eqs.~\eqref{mu} and~\eqref{nu}) does not invalidate the null test since it is factorized out in Eq.~\eqref{N0vanish}. 

This illustrates that using a $\Lambda$CDM model to calculate the weights is in no sense a restrictive assumption. It is just a convenient choice, which leads to a vanishing $\NN$ whenever relation~\eqref{eq:Pevol} holds, i.e.\ whenever the growth of structure is scale-independent~\footnote{Note that instead of calculating the weights $\mu_2, \mu_4$ and $\nu_1$ with a $\Lambda$CDM power spectrum, we could measure the monopole of the power spectrum $P_0(k,z)$ and calculate the weights with it. We have however tested that the uncertainty in the measurement of $P_0(k,z)$ degrades the precision of the null test and that it is therefore more efficient to use $\bar P(k,z)$. }.

\section{Results}

\begin{figure}
\centering
\includegraphics[width=0.9\columnwidth]{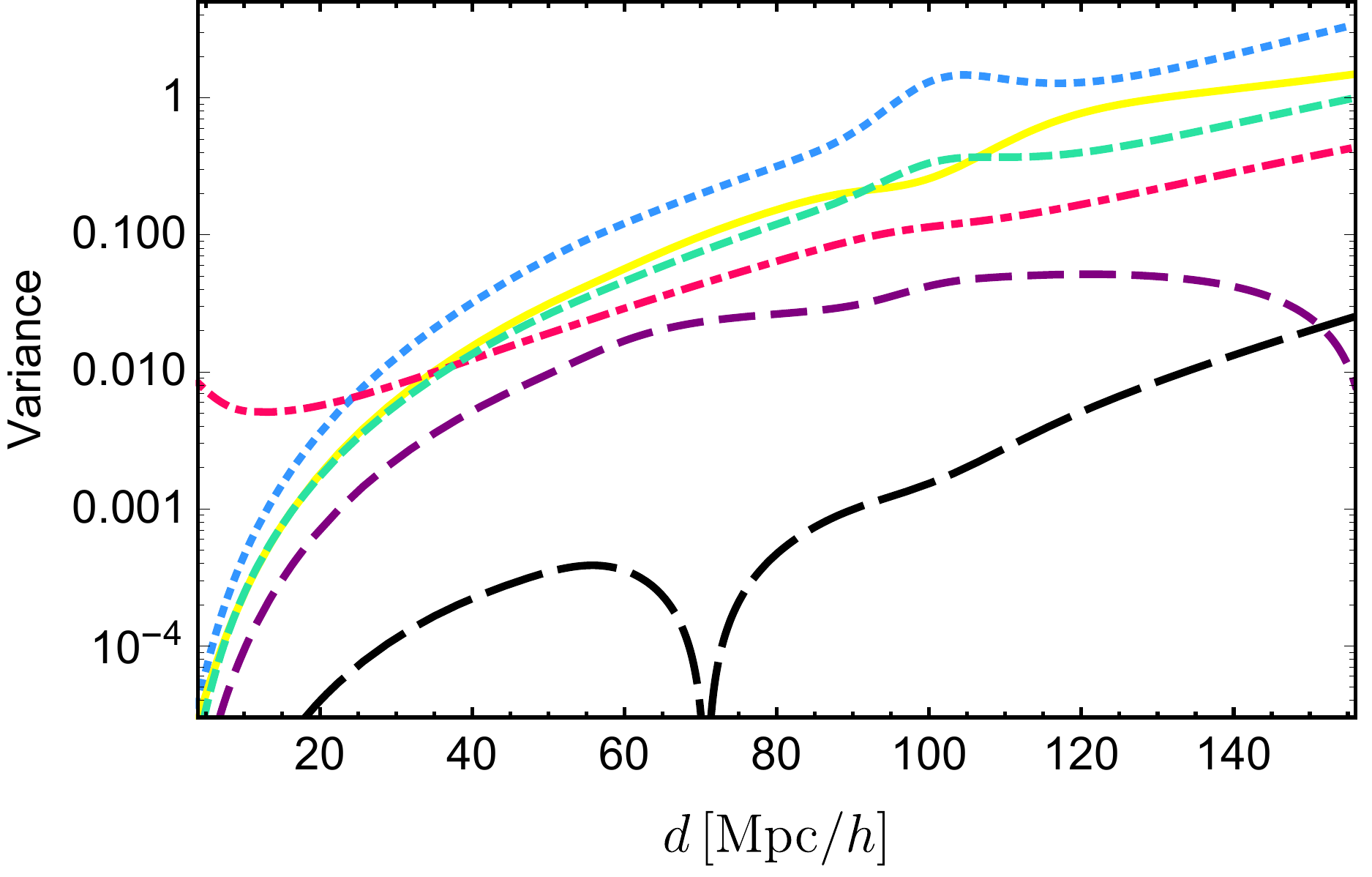}
\caption{Variance for a survey like DESI, plotted as a function of separation at $z=0.15$. We show the contributions from the: quadrupole (blue dotted), hexadecapole (yellow solid), dipole (red dot-dashed), dipole-quadrupole (green short-dashed), quadrupole-hexadecapole (purple middle-dashed) and dipole-hexadecapole (black long-dashed).}
\label{fig:variance}
\end{figure}

The sensitivity of the null test to the scale-dependence of $f$ is determined by its covariance. Since $\hNN$ is a sum of multipoles, its covariance is due to the variance of each multipole, plus the covariance between them. We follow the method developed in~\cite{Hall:2016bmm,Tansella:2018sld} to calculate each of these terms.
We have contributions from the cosmic variance of $\Delta$, of $\kappa$ and the covariance between them. In addition, we have a contribution from shot noise, which affects $\Delta$; and a contribution from the error in the determination of $\kappa$. \cite{2012ApJ...744L..22S,Casaponsa:2012tq,Heavens:2013gol,Alsing:2014fya} proposed an estimator to measure $\kappa$ by combining the size and luminosity of galaxies. In the forecast we consider a spectroscopic survey with specifications like the Bright Galaxy Sample (BCG) of DESI~\citep{Aghamousa:2016zmz}, that will measure the redshift, position and luminosity of 10 million of galaxies. Sizes will be measured by the DESI Legacy Imaging Survey~\citep{Dey_2019}. The error in the measurement of the size depends on the resolution of the instrument but also on the type of galaxies. As discussed in~\cite{Alsing:2014fya}, the sizes of late-type (spiral) galaxies tend to be better measured than for early-type (elliptical) galaxies. This results in an error on $\kappa$ ranging from $\sigma_\kappa=0.3$ to $\sigma_\kappa=0.8$. In the following we use $\sigma_\kappa=0.3$ as fiducial value, and we explore how the constraints degrade when $\sigma_\kappa=0.8$.
In Fig.~\ref{fig:variance} we show the different contributions to the variance. We see that at small separations, the dominant contribution is due to the dipole, more particularly to the error in the measurement of the convergence $\sigma_\kappa$. At large separations on the other hand, the dominant contribution is due to the cosmic variance of the quadrupole.

\begin{figure}
\centering
\includegraphics[width=0.9\columnwidth]{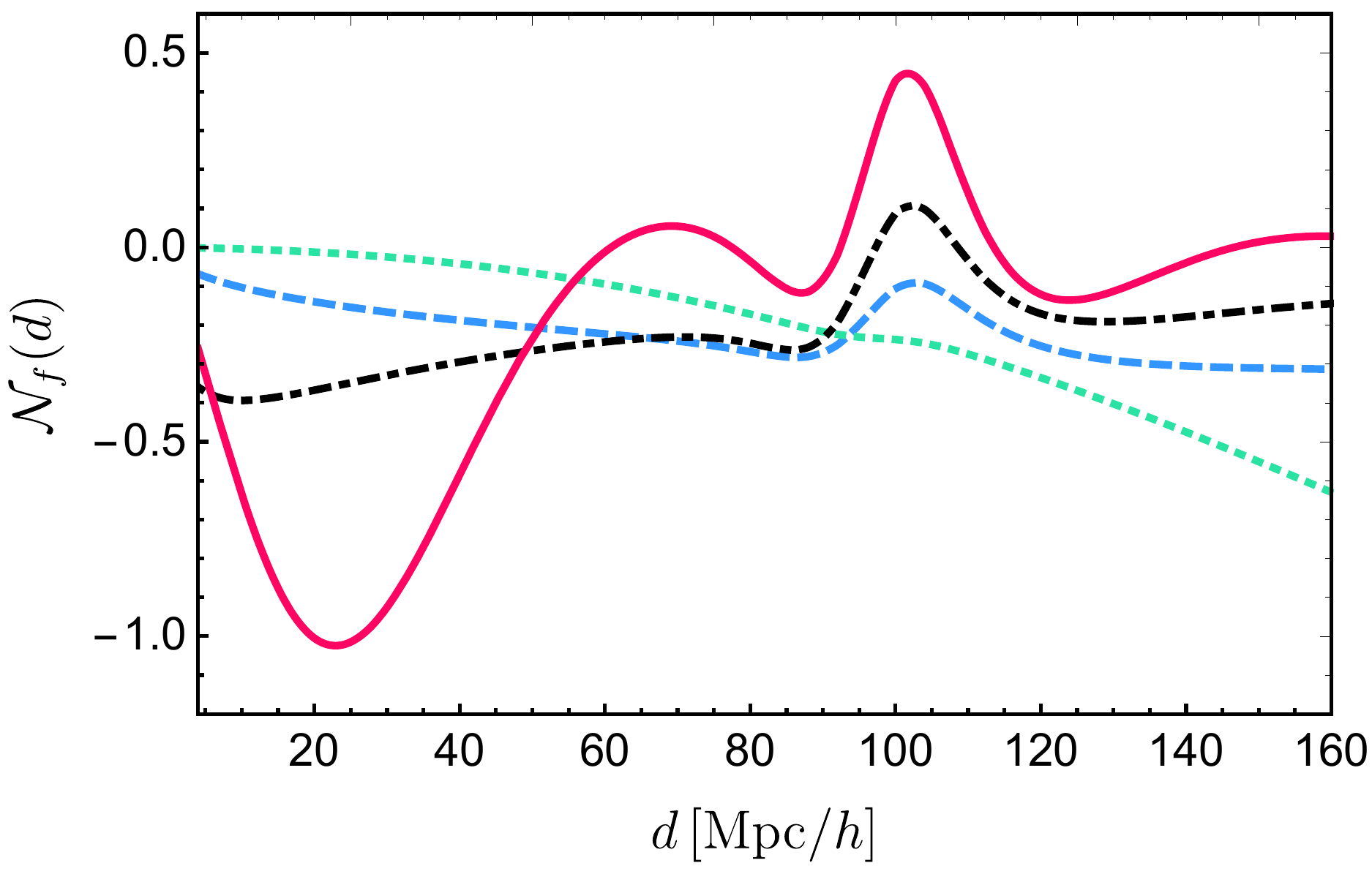}\\
\includegraphics[width=0.9\columnwidth]{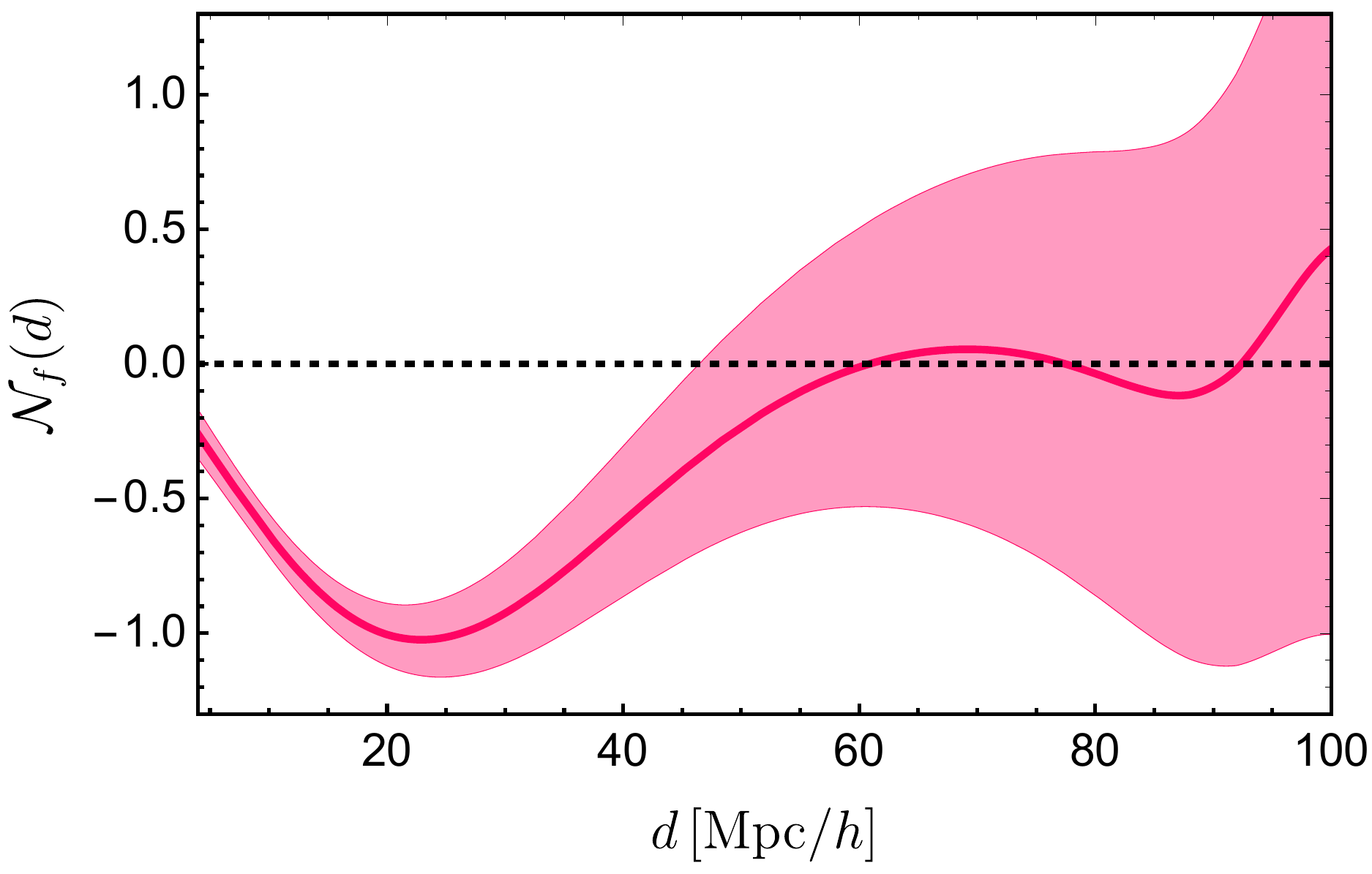}
\caption{\emph{Top}: $\NN$ plotted as a function of separation at $z=0.15$, for four different models with $\epsilon_0=0.5$, $c_1=1$, $c_2=0$, and with: $\ks=0.01\,h$/Mpc, $m=-1$ (blue dashed), $\ks=0.01\,h$/Mpc, $m=-4$ (green dotted), $\ks=0.1\,h$/Mpc, $m=-1$ (black dot-dashed) and $\ks=0.1\,h$/Mpc, $m=-4$ (red solid). 
\emph{Bottom}: $\NN$ for the last model, plotted with its variance for a survey with DESI specifications.}
\label{fig:N0}
\end{figure}

To apply the null test on data we do not need any modelling of the growth of structure: we simply combine the measured $\hxiq, \hxih$ and $\hxid$ according to Eq.~\eqref{N0def} and see if the resulting $\hNN$ is consistent or not with zero. Since DESI data are however not yet available, we want to forecast how sensitive the null test is expected to be to a given scale-dependence of $f$. For this we choose a parameterization of $D_1$ and we forecast how sensitive the null test is. We take 
\begin{align}
D_1(z,k)=\bar D_1(z)\Big[1+\epsilon(z)\gamma(k) \Big]\,,\ 
\gamma(k)=c_1\frac{1+c_2 (k/\ks)^\nk}{1+(k/\ks)^\nk}\, . \label{eq:gamma}
\end{align}
The coefficients $c_1$ and $c_2$ govern the amplitude of $\gamma$ for large and small scales, $\ks$ determines the scale of the transition from one regime to the other, and $m$ its slope. The amplitude of the deviations is encoded in $\epsilon(z)$, so we choose $c_1$ and $c_2$ such that $0\leq \gamma(k)\leq 1$. We assume that the evolution of $\epsilon(z)$ follows that of dark energy, so that it becomes negligible in the past:
$
\epsilon(z)=\epsilon_0\frac{\Omega_\Lambda(z)}{\Omega_{\Lambda}(z=0)}\, ,
$
where $\Omega_\Lambda(z)$ is the density parameter of the cosmological constant, and $\epsilon_0$ is a free parameter. In Fig.~\ref{fig:N0}, we plot $\NN$  for four different models. When $\ks=0.1\,h$/Mpc, the deviations are more important at small separations, whereas for $\ks=0.01\,h$/Mpc they increase at large separations. The slope $m$ also has a significant impact on the form of $\NN$. For comparison the $f(R)$ model explored in~\cite{Giannantonio:2009gi} has $\ks\sim 0.05\,h$/Mpc.

\begin{table}
\begin{center}
\begin{tabular}{|c|C{0.7cm}C{0.7cm}C{0.7cm}|C{0.7cm}C{0.7cm}C{0.7cm}|}
\hline
  & \multicolumn{3}{c|}{$\ks=0.1\,h/$Mpc} & \multicolumn{3}{c|}{$\ks=0.01\,h/$Mpc}  \\
\hline
 $d_{\rm min}$ & \multicolumn{3}{c|}{$m$} & \multicolumn{3}{c|}{$m$}  \\
 $[$Mpc/$h]$& -1 & -2 & -4 & -1 & -2 & -4  \\
\hline
20& 0.23&0.13&0.08&0.52&0.88&1.60\\
32&0.37&0.21&0.13&0.65&0.92&1.60\\
40& 0.47& 0.30& 0.21& 0.74 & 0.96& 1.61\\
\hline
\end{tabular}
\end{center}
\caption{Minimal values of $\epsilon_0$ leading to a measurement of $\NN$ different from 0 with a significance of $3\sigma$, for a survey like DESI. We show 6 models 
 with $c_1=1, c_2=0$, and $\ks$, $m$, and $d_{\rm min}$ shown. We fix $d_{\rm max}=156$\,Mpc/$h$; the redshift range is $0\leq z\leq 0.5$.}
\label{tab:DESI}

\end{table}

To assess the sensitivity of the null test to scale-dependence, we forecast the constraints that can be obtained on $\epsilon_0$ for some fixed representative choices of the parameters $c_1, c_2,\nk$ and $\ks$.
We do not marginalize over these parameters, because our aim is not to fit a certain model. We rather want to determine how sensitive the null test is to a generic scale-dependence. 
We fix the cosmological parameters to their fiducial value taken from~\cite{Aghanim:2018eyx}, neglecting massive neutrinos
and we construct the Fisher matrix 
\be
\FF_{\epsilon_0}=\sum_{i,j,z}\frac{\partial \NN(d_i, z)}{\partial \epsilon_0}\big[{\rm cov}(\hNN)\big]^{-1}(d_i, d_j, z)\frac{\partial \NN(d_j, z)}{\partial \epsilon_0}\, , \nonumber
\ee
summing over the redshift bins and the pixels separations 
between $d_{\rm min}$ and $d_{\rm max}$. The minimum separation $d_{\rm min}$ is determined by the scale at which non-linearities invalidate the null test. In Section~\ref{sec:contaminations} we show that below $\sim$ 30 Mpc/$h$, non-linearities generate a $\NN$ which is larger that the variance. Above this scale however, linear perturbation theory is accurate enough. In the forecasts we choose 3 values for $d_{\rm min}$: 20, 32 and 40 Mpc/$h$.
The results for a survey like the BCG of DESI are summarized in Table~\ref{tab:DESI}. The constraints are significantly better for the models with $\ks=0.1\,h$/Mpc, since in this case the deviations in $\NN$ are important at smaller scales, where cosmic variance is smaller. The constraints  degrade increasing $d_{\rm min}$, but even with a large $d_{\rm min}=40$\,Mpc/$h$, DESI is sensitive to deviations $\sim$20\%. They decrease to 8\% for $d_{\rm min}=20$\,Mpc/$h$. 
The constraints are also sensitive to the precision in the size measurements. Increasing $\sigma_\kappa$ from 0.3 to 0.8, we degrade the constraints by a factor 1.5-2. On the other hand, increasing the number density and volume to the ones planed for SKA phase 2~\citep{Bull:2015lja}, the constraints are improved by a factor 2.

Comparing with current constraints on the growth rate $f$ in specific models, an $f(R)$ model with $|f_{R0}|=3.2\times 10^{-5}$ leads to a scale-dependence of 20\% in $f$ in the range $k\in [10^{-3}-10^{-1}]\,$Mpc$^{-1}$, whereas current RSD constraints give $|f_{R0}|< 10^{-4}$~\citep{Song:2015oza}.
Planck constraints on a generic scale-dependent $\mu(z,k)$ are of order 1~\citep{Ade:2015rim}.
In comparison, our third model in Table~\ref{tab:DESI} ($d_{\rm min}=20$\,Mpc/$h$), generates a scale-dependence of 10\% in $f$.

\begin{figure}
\centering
\includegraphics[width=0.9\columnwidth]{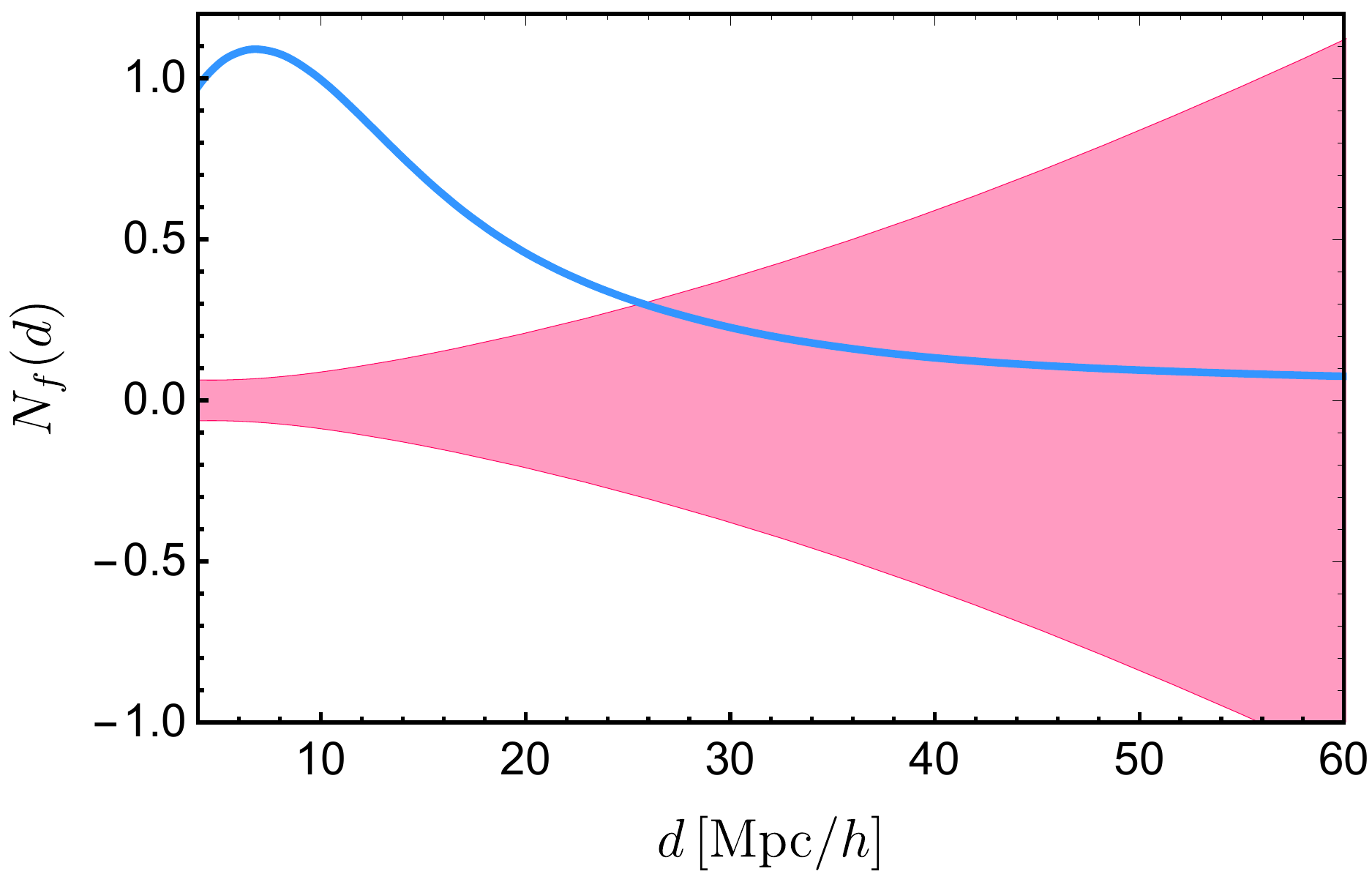}
\caption{$\NN$ at $z=0.05$ calculated with the streaming model for RSD, with parameters from
~\citet{Xu_2013}: $\Sigma_s=4$\,Mpc/$h$, $\Sigma_\parallel=10$\,Mpc/$h$ and $\Sigma_\perp=6$\,Mpc/$h$ (blue line). The red region shows the variance of $\NN$ for a survey with DESI specifications.}
\label{fig:NL}
\end{figure}

\begin{figure}
\centering
\includegraphics[width=0.9\columnwidth]{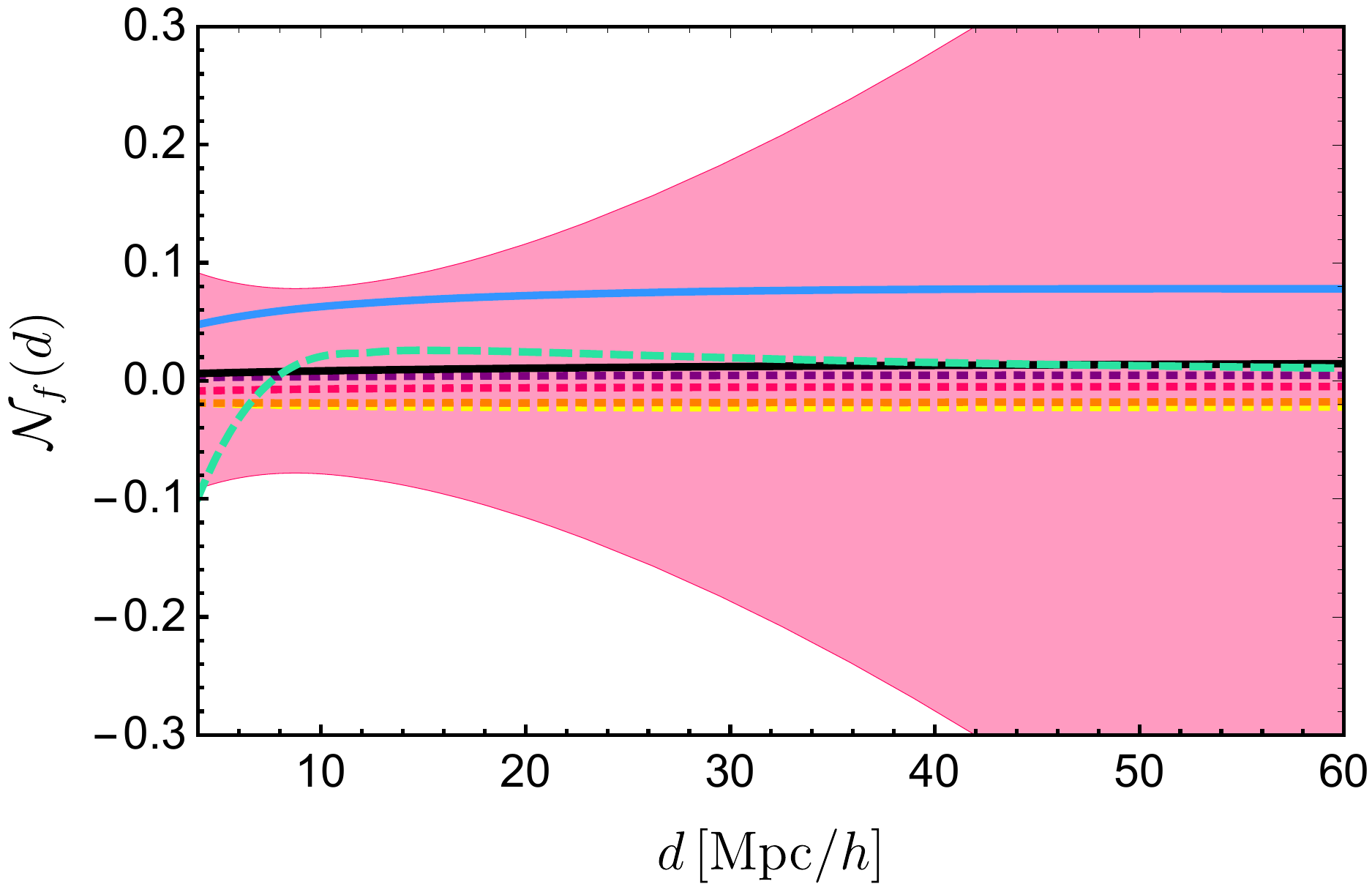}
\caption{Value of $\NN$ at $z=0.15$ obtained by varying the fiducial cosmology by 2$\sigma$: $\Omega_m$ (yellow dotted), $\Omega_b$ (purple dotted), $n_s$ (red dotted) and $h$ (orange dotted). The green dashed line shows $\NN$ with a scale-dependent bias. The black solid line and blue solid line show $\NN$ for a cosmology with massive neutrinos with mass $m_\nu=0.12$\,eV (black) and $m_\nu=0.6$\,eV (blue). The red region shows the variance of $\NN$ for a survey with DESI specifications.}
\label{fig:cont}
\end{figure}

\section{Contaminations}
\label{sec:contaminations}

We now explore the limitations of the null test, i.e.\ the situations where $\NN\neq 0$ even if $D_1$ and $f$ are scale-independent. First we assess the impact of non-linearities on $\NN$. We use the streaming model for RSD that has been used to analyse SDSS data~\citep{Xu_2013}, which contains both the Fingers of God effect~\citep{Peacock:1993xg} and the impact of non-linearities on the BAO~\citep{Eisenstein:2006nj}. We extend this model to the dipole, assuming that non-linearities in the velocity affect the dipole in the same way as they affect the multipoles of galaxy clustering. In Fig.~\ref{fig:NL}, we show $\NN$ obtained by using the streaming model for $\xiq, \xih$ and $\xid$, and the linear power spectrum for the coefficients $\bar\mu_\ell$ and $\bar\nu_1$. We see that below $\sim$ 30\,Mpc/$h$, non-linearities generate an $\NN$ which is significantly larger than the variance. Above this scale however, non-linearities become smaller than the variance, meaning that they do not limit the validity and sensitivity of the null test.

Second, we assess the impact of the fiducial cosmology by calculating $\NN$ with coefficients computed in the fiducial cosmology, and observables 2$\sigma$ away from fiducial (we take $\sigma$ from Table 2 of~\cite{Aghanim:2018eyx}). In Fig.~\ref{fig:cont}, we compare $\NN$ obtained in this way, with its variance for a survey with DESI specifications. Varying the cosmology generates a $\NN$ which is significantly smaller than the variance. This means that a detection of $\NN\neq 0$ cannot be due to our choice of fiducial cosmology, except if Planck constraints are incorrect by more than 2$\sigma$.
Let us furthermore mention that we can always test this assumption by fitting the null test to 0 varying the cosmological parameters. Any remaining non-zero $\NN$ will then be purely due to a scale-dependent growth rate. 

The third limitation comes from a possible scale-dependent bias, that would induce a non-zero $\NN$. In Fig.~\ref{fig:cont}, we show $\NN$ obtained for a particular choice of scale-dependent bias used in~\cite{Amendola:2015pha} and fitted from simulations:
$
b(z,k)=b_0(z)\sqrt{1+Q(z)\left(k/k_1\right)^2}/\sqrt{1+A(z)k/k_1}\, ,
$
with $A(z)=1.7$, $Q(z)$ fitted from~\cite{Amendola:2015pha}, $k_1=1\,h/$Mpc and $b_0(z)$ from DESI. We see that below 10\,Mpc/$h$ the scale-dependent bias induces a $\NN$ which is of the same order as the variance. A detection at those scales could therefore be due to the bias. At larger scales however, the scale-dependent bias has a negligible impact, meaning that the null test is robust above 10\,Mpc/$h$. 

Massive neutrinos also lead to a scale-dependent growth of structure~\citep{Lesgourgues:2006nd}. In Fig.~\ref{fig:cont}, we show $\NN$ induced by a cosmology with two massless neutrinos and one massive neutrinos, for two choices of mass. We use CAMB to compute the density and velocity transfer functions with massive neutrinos~\citep{Lewis:1999bs,2012JCAP...04..027H}. We see that neutrinos become relevant (i.e.\ $\NN$ becomes of the same order as the variance) only for a large mass of $\sim 0.6$\,eV. This is consistent with RSD forecasts that are sensitive to neutrinos masses of this order~\citep{2011MNRAS.418..346M}.

Another possible source of contamination are wide-angle effects. These effects have been shown to be an important contamination to the dipole of galaxy clustering, especially when the line-of-sight used to compute the multipoles break the symmetry of the configuration~\citep{Beutler:2018vpe,Gaztanaga:2015jrs}. To minimise this effect we use as line-of-sight the median to the galaxy pair. With this, we compute $\NN$ using the full-sky expression for the multipoles $\xiq, \xih$ and $\xid$. In the lowest redshift bin of DESI, $z\leq 0.05$, we see that wide-angle effects invalidate the null test above 60\,Mpc/$h$. For all other redshift bins however, they are negligible over the whole range of scales we are using. 

Finally, lensing can also potentially contaminate the null test since it affects both the galaxy number counts $\Delta$ and the convergence $\kappa$. Including these two contributions in $\NN$ we find that for the BCG sample of DESI, which is limited to $z\leq0.5$, lensing is always subdominant and does not impact our results. If one wants to use the null test at higher redshift however, this contribution would have to be included as a contamination, which would limit the sensitivity of the null test.

\section{Conclusion}

In this letter we have constructed a null test to probe the scale-dependence of the growth of structure. The standard way of testing for such a scale dependence is to compare observables with theoretical predictions for a specific modelling of $f(k,z)$ and to measure the parameters of this model. The drawback of this method is that if the modelling is incorrect, it can invalidate the test. One needs therefore to test one after the other a large number of models. The null test presented in this letter has the advantage of being model-independent: to apply it on data, one simply has to combine observables as prescribed in Eq.~\eqref{N0def} and see if the result is consistent or not with zero. A non-zero $\NN$ can then directly be interpreted as a deviation from scale-dependence.
An ideal null test should not depend on any assumption about cosmology. Here we have shown that this is not possible, since we need a fiducial cosmology to calculate the coefficients of the null test. However, we have demonstrated that the assumptions that we use are very general and have little impact on the validity of the null test. In short, the null test is valid as long as the background evolution of the Universe at late times is consistent with $\Lambda$CDM, and that the cosmological parameters $\Omega_b$, $\Omega_m$, $n_s$ and $h$ are consistent with Planck constraints. Under these assumptions, the null test vanishes for \emph{any} form of the growth rate $f$ which is scale-independent. The price to pay for this generality is that the null test is limited by the covariance of all observables, which in total is larger than the covariance of individual observables. Hence, for a specific model, the null test will perform worse than individual observables. This test should therefore be used as a first \emph{model-independent}  discriminating method between scale-dependent and independent models.
We have seen that the null test will be sensitive to deviations of the order of 10-20 percent for DESI, and of 5-10 percent for SKA2, making it a very valuable and powerful tool for upcoming surveys.

\paragraph*{Acknowledgements}
We thank David Bacon, Ruth Durrer and Martin Kunz for useful discussions and Goran Jelic-Cizmek for his help with COFFE. F.O.F.\ and C.B.\ acknowledge support by the Swiss National Science Foundation. C.C.\ was supported by STFC Consolidated Grant ST/P000592/1.

\bibliographystyle{mnras}
\bibliography{null_test.bib}

\begin{thebibliography}{}
\makeatletter
\relax
\def\mn@urlcharsother{\let\do\@makeother \do\$\do\&\do\#\do\^\do\_\do\%\do\~}
\def\mn@doi{\begingroup\mn@urlcharsother \@ifnextchar [ {\mn@doi@}
  {\mn@doi@[]}}
\def\mn@doi@[#1]#2{\def\@tempa{#1}\ifx\@tempa\@empty \href
  {http://dx.doi.org/#2} {doi:#2}\else \href {http://dx.doi.org/#2} {#1}\fi
  \endgroup}
\def\mn@eprint#1#2{\mn@eprint@#1:#2::\@nil}
\def\mn@eprint@arXiv#1{\href {http://arxiv.org/abs/#1} {{\tt arXiv:#1}}}
\def\mn@eprint@dblp#1{\href {http://dblp.uni-trier.de/rec/bibtex/#1.xml}
  {dblp:#1}}
\def\mn@eprint@#1:#2:#3:#4\@nil{\def\@tempa {#1}\def\@tempb {#2}\def\@tempc
  {#3}\ifx \@tempc \@empty \let \@tempc \@tempb \let \@tempb \@tempa \fi \ifx
  \@tempb \@empty \def\@tempb {arXiv}\fi \@ifundefined
  {mn@eprint@\@tempb}{\@tempb:\@tempc}{\expandafter \expandafter \csname
  mn@eprint@\@tempb\endcsname \expandafter{\@tempc}}}

\bibitem[\protect\citeauthoryear{Aghamousa et~al.}{Aghamousa
  et~al.}{2016}]{Aghamousa:2016zmz}
Aghamousa A.,  et~al., 2016

\bibitem[\protect\citeauthoryear{Aghanim et~al.}{Aghanim
  et~al.}{2018}]{Aghanim:2018eyx}
Aghanim N.,  et~al., 2018

\bibitem[\protect\citeauthoryear{Alsing, Kirk, Heavens  \& Jaffe}{Alsing
  et~al.}{2015}]{Alsing:2014fya}
Alsing J.,  Kirk D.,  Heavens A.,   Jaffe A.,  2015, \mn@doi [Mon. Not. Roy.
  Astron. Soc.] {10.1093/mnras/stv1249}, 452, 1202

\bibitem[\protect\citeauthoryear{Amendola, Menegoni, Di~Porto, Corsi  \&
  Branchini}{Amendola et~al.}{2017}]{Amendola:2015pha}
Amendola L.,  Menegoni E.,  Di~Porto C.,  Corsi M.,   Branchini E.,  2017,
  \mn@doi [Phys. Rev.] {10.1103/PhysRevD.95.023505}, D95, 023505

\bibitem[\protect\citeauthoryear{Bacon, Andrianomena, Clarkson, Bolejko  \&
  Maartens}{Bacon et~al.}{2014}]{Bacon:2014uja}
Bacon D.~J.,  Andrianomena S.,  Clarkson C.,  Bolejko K.,   Maartens R.,  2014,
  \mn@doi [Mon. Not. Roy. Astron. Soc.] {10.1093/mnras/stu1270}, 443, 1900

\bibitem[\protect\citeauthoryear{Baker, Ferreira  \& Skordis}{Baker
  et~al.}{2013}]{Baker:2012zs}
Baker T.,  Ferreira P.~G.,   Skordis C.,  2013, \mn@doi [Phys. Rev.]
  {10.1103/PhysRevD.87.024015}, D87, 024015

\bibitem[\protect\citeauthoryear{Beutler, Castorina  \& Zhang}{Beutler
  et~al.}{2019}]{Beutler:2018vpe}
Beutler F.,  Castorina E.,   Zhang P.,  2019, \mn@doi [JCAP]
  {10.1088/1475-7516/2019/03/040}, 1903, 040

\bibitem[\protect\citeauthoryear{Bolejko, Clarkson, Maartens, Bacon, Meures  \&
  Beynon}{Bolejko et~al.}{2013}]{Bolejko:2012uj}
Bolejko K.,  Clarkson C.,  Maartens R.,  Bacon D.,  Meures N.,   Beynon E.,
  2013, \mn@doi [Phys. Rev. Lett.] {10.1103/PhysRevLett.110.021302}, 110,
  021302

\bibitem[\protect\citeauthoryear{Bonvin}{Bonvin}{2008}]{Bonvin:2008ni}
Bonvin C.,  2008, \mn@doi [Phys. Rev.] {10.1103/PhysRevD.78.123530}, D78,
  123530

\bibitem[\protect\citeauthoryear{Bonvin \& Durrer}{Bonvin \&
  Durrer}{2011}]{Bonvin:2011bg}
Bonvin C.,  Durrer R.,  2011, \mn@doi [Phys. Rev.]
  {10.1103/PhysRevD.84.063505}, D84, 063505

\bibitem[\protect\citeauthoryear{Bonvin, Andrianomena, Bacon, Clarkson,
  Maartens, Moloi  \& Bull}{Bonvin et~al.}{2017}]{Bonvin:2016dze}
Bonvin C.,  Andrianomena S.,  Bacon D.,  Clarkson C.,  Maartens R.,  Moloi T.,
   Bull P.,  2017, \mn@doi [Mon. Not. Roy. Astron. Soc.]
  {10.1093/mnras/stx2049}, 472, 3936

\bibitem[\protect\citeauthoryear{{Buchdahl}}{{Buchdahl}}{1970}]{1970MNRAS.150....1B}
{Buchdahl} H.~A.,  1970, \mn@doi [\mnras] {10.1093/mnras/150.1.1}, \href
  {https://ui.adsabs.harvard.edu/abs/1970MNRAS.150....1B} {150, 1}

\bibitem[\protect\citeauthoryear{Bull}{Bull}{2016}]{Bull:2015lja}
Bull P.,  2016, \mn@doi [Astrophys. J.] {10.3847/0004-637X/817/1/26}, 817, 26

\bibitem[\protect\citeauthoryear{Casaponsa, Heavens, Kitching, Miller, Barreiro
   \& Martinez-Gonzalez}{Casaponsa et~al.}{2013}]{Casaponsa:2012tq}
Casaponsa B.,  Heavens A.~F.,  Kitching T.~D.,  Miller L.,  Barreiro R.~B.,
  Martinez-Gonzalez E.,  2013, \mn@doi [Mon. Not. Roy. Astron. Soc.]
  {10.1093/mnras/stt088}, 430, 2844

\bibitem[\protect\citeauthoryear{Challinor \& Lewis}{Challinor \&
  Lewis}{2011}]{Challinor:2011bk}
Challinor A.,  Lewis A.,  2011, \mn@doi [Phys. Rev.]
  {10.1103/PhysRevD.84.043516}, D84, 043516

\bibitem[\protect\citeauthoryear{De~Felice, Kobayashi  \& Tsujikawa}{De~Felice
  et~al.}{2011}]{DeFelice:2011hq}
De~Felice A.,  Kobayashi T.,   Tsujikawa S.,  2011, \mn@doi [Phys. Lett.]
  {10.1016/j.physletb.2011.11.028}, B706, 123

\bibitem[\protect\citeauthoryear{Dey et~al.,}{Dey et~al.}{2019}]{Dey_2019}
Dey A.,  et~al., 2019, \mn@doi [The Astronomical Journal]
  {10.3847/1538-3881/ab089d}, 157, 168

\bibitem[\protect\citeauthoryear{Dodelson}{Dodelson}{2003}]{Dodelson:2003ft}
Dodelson S.,  2003, {Modern Cosmology}.
Academic Press, Amsterdam

\bibitem[\protect\citeauthoryear{Eisenstein, Seo  \& White}{Eisenstein
  et~al.}{2007}]{Eisenstein:2006nj}
Eisenstein D.~J.,  Seo H.-j.,   White M.~J.,  2007, \mn@doi [Astrophys. J.]
  {10.1086/518755}, 664, 660

\bibitem[\protect\citeauthoryear{Gaztanaga, Bonvin  \& Hui}{Gaztanaga
  et~al.}{2017}]{Gaztanaga:2015jrs}
Gaztanaga E.,  Bonvin C.,   Hui L.,  2017, \mn@doi [JCAP]
  {10.1088/1475-7516/2017/01/032}, 1701, 032

\bibitem[\protect\citeauthoryear{Ghosh \& Durrer}{Ghosh \&
  Durrer}{2019}]{Ghosh:2018ijm}
Ghosh B.,  Durrer R.,  2019, \mn@doi [JCAP] {10.1088/1475-7516/2019/06/010},
  1906, 010

\bibitem[\protect\citeauthoryear{Giannantonio, Martinelli, Silvestri  \&
  Melchiorri}{Giannantonio et~al.}{2010}]{Giannantonio:2009gi}
Giannantonio T.,  Martinelli M.,  Silvestri A.,   Melchiorri A.,  2010, \mn@doi
  [JCAP] {10.1088/1475-7516/2010/04/030}, 1004, 030

\bibitem[\protect\citeauthoryear{Gleyzes, Langlois, Mancarella  \&
  Vernizzi}{Gleyzes et~al.}{2016}]{Gleyzes:2015rua}
Gleyzes J.,  Langlois D.,  Mancarella M.,   Vernizzi F.,  2016, \mn@doi [JCAP]
  {10.1088/1475-7516/2016/02/056}, 1602, 056

\bibitem[\protect\citeauthoryear{Gubitosi, Piazza  \& Vernizzi}{Gubitosi
  et~al.}{2013}]{Gubitosi:2012hu}
Gubitosi G.,  Piazza F.,   Vernizzi F.,  2013, \mn@doi [JCAP]
  {10.1088/1475-7516/2013/02/032}, 1302, 032

\bibitem[\protect\citeauthoryear{Hall \& Bonvin}{Hall \&
  Bonvin}{2017}]{Hall:2016bmm}
Hall A.,  Bonvin C.,  2017, \mn@doi [Phys. Rev.] {10.1103/PhysRevD.95.043530},
  D95, 043530

\bibitem[\protect\citeauthoryear{Hamilton}{Hamilton}{1997}]{Hamilton:1997zq}
Hamilton A. J.~S.,  1997, in {Ringberg Workshop on Large Scale Structure
  Ringberg, Germany, September 23-28, 1996}.  (\mn@eprint {arXiv}
  {astro-ph/9708102}), \mn@doi{10.1007/978-94-011-4960-0_17}

\bibitem[\protect\citeauthoryear{Heavens, Alsing  \& Jaffe}{Heavens
  et~al.}{2013}]{Heavens:2013gol}
Heavens A.,  Alsing J.,   Jaffe A.,  2013, \mn@doi [Mon. Not. Roy. Astron.
  Soc.] {10.1093/mnrasl/slt045}, 433, 6

\bibitem[\protect\citeauthoryear{{Howlett}, {Lewis}, {Hall}  \&
  {Challinor}}{{Howlett} et~al.}{2012}]{2012JCAP...04..027H}
{Howlett} C.,  {Lewis} A.,  {Hall} A.,   {Challinor} A.,  2012, \mn@doi
  [Journal of Cosmology and Astro-Particle Physics]
  {10.1088/1475-7516/2012/04/027}, \href
  {https://ui.adsabs.harvard.edu/abs/2012JCAP...04..027H} {2012, 027}

\bibitem[\protect\citeauthoryear{Jeong, Schmidt  \& Hirata}{Jeong
  et~al.}{2012}]{Jeong:2011as}
Jeong D.,  Schmidt F.,   Hirata C.~M.,  2012, \mn@doi [Phys. Rev.]
  {10.1103/PhysRevD.85.023504}, D85, 023504

\bibitem[\protect\citeauthoryear{Kaiser}{Kaiser}{1987}]{Kaiser:1987qv}
Kaiser N.,  1987, Mon. Not. Roy. Astron. Soc., 227, 1

\bibitem[\protect\citeauthoryear{Lesgourgues \& Pastor}{Lesgourgues \&
  Pastor}{2006}]{Lesgourgues:2006nd}
Lesgourgues J.,  Pastor S.,  2006, \mn@doi [Phys. Rept.]
  {10.1016/j.physrep.2006.04.001}, 429, 307

\bibitem[\protect\citeauthoryear{Lewis, Challinor  \& Lasenby}{Lewis
  et~al.}{2000}]{Lewis:1999bs}
Lewis A.,  Challinor A.,   Lasenby A.,  2000, \mn@doi [Astrophys. J.]
  {10.1086/309179}, 538, 473

\bibitem[\protect\citeauthoryear{Linder}{Linder}{2005}]{PhysRevD.72.043529}
Linder E.~V.,  2005, \mn@doi [Phys. Rev. D] {10.1103/PhysRevD.72.043529}, 72,
  043529

\bibitem[\protect\citeauthoryear{{Marulli}, {Carbone}, {Viel}, {Moscardini}  \&
  {Cimatti}}{{Marulli} et~al.}{2011}]{2011MNRAS.418..346M}
{Marulli} F.,  {Carbone} C.,  {Viel} M.,  {Moscardini} L.,   {Cimatti} A.,
  2011, \mn@doi [Mon. Not. Roy. Astron. Soc.]
  {10.1111/j.1365-2966.2011.19488.x}, \href
  {https://ui.adsabs.harvard.edu/abs/2011MNRAS.418..346M} {418, 346}

\bibitem[\protect\citeauthoryear{Peacock \& Dodds}{Peacock \&
  Dodds}{1994}]{Peacock:1993xg}
Peacock J.~A.,  Dodds S.~J.,  1994, \mn@doi [Mon. Not. Roy. Astron. Soc.]
  {10.1093/mnras/267.4.1020}, 267, 1020

\bibitem[\protect\citeauthoryear{{{Planck Collaboration XIV}}}{{{Planck
  Collaboration XIV}}}{2016}]{Ade:2015rim}
{{Planck Collaboration XIV}} 2016, \mn@doi [Astron. Astrophys.]
  {10.1051/0004-6361/201525814}, 594, A14

\bibitem[\protect\citeauthoryear{Sawicki \& Bellini}{Sawicki \&
  Bellini}{2015}]{Sawicki:2015zya}
Sawicki I.,  Bellini E.,  2015, \mn@doi [Phys. Rev.]
  {10.1103/PhysRevD.92.084061}, D92, 084061

\bibitem[\protect\citeauthoryear{Schmidt, Leauthaud, Massey, Rhodes, George,
  Koekemoer, Finoguenov  \& Tanaka}{Schmidt et~al.}{2012a}]{Schmidt:2011qj}
Schmidt F.,  Leauthaud A.,  Massey R.,  Rhodes J.,  George M.~R.,  Koekemoer
  A.~M.,  Finoguenov A.,   Tanaka M.,  2012a, \mn@doi [Astrophys. J.]
  {10.1088/2041-8205/744/2/L22}, 744, L22

\bibitem[\protect\citeauthoryear{{Schmidt}, {Leauthaud}, {Massey}, {Rhodes},
  {George}, {Koekemoer}, {Finoguenov}  \& {Tanaka}}{{Schmidt}
  et~al.}{2012b}]{2012ApJ...744L..22S}
{Schmidt} F.,  {Leauthaud} A.,  {Massey} R.,  {Rhodes} J.,  {George} M.~R.,
  {Koekemoer} A.~M.,  {Finoguenov} A.,   {Tanaka} M.,  2012b, \mn@doi [ApJL]
  {10.1088/2041-8205/744/2/L22}, \href
  {http://adsabs.harvard.edu/abs/2012ApJ...744L..22S} {744, L22}

\bibitem[\protect\citeauthoryear{Song et~al.,}{Song
  et~al.}{2015}]{Song:2015oza}
Song Y.-S.,  et~al., 2015, \mn@doi [Phys. Rev.] {10.1103/PhysRevD.92.043522},
  D92, 043522

\bibitem[\protect\citeauthoryear{{Starobinsky}}{{Starobinsky}}{1980}]{1980PhLB...91...99S}
{Starobinsky} A.~A.,  1980, \mn@doi [Physics Letters B]
  {10.1016/0370-2693(80)90670-X}, \href
  {https://ui.adsabs.harvard.edu/abs/1980PhLB...91...99S} {91, 99}

\bibitem[\protect\citeauthoryear{Tansella, Jelic-Cizmek, Bonvin  \&
  Durrer}{Tansella et~al.}{2018}]{Tansella:2018sld}
Tansella V.,  Jelic-Cizmek G.,  Bonvin C.,   Durrer R.,  2018, \mn@doi [JCAP]
  {10.1088/1475-7516/2018/10/032}, 1810, 032

\bibitem[\protect\citeauthoryear{Wang \& Steinhardt}{Wang \&
  Steinhardt}{1998}]{Wang_1998}
Wang L.,  Steinhardt P.~J.,  1998, \mn@doi [The Astrophysical Journal]
  {10.1086/306436}, 508, 483

\bibitem[\protect\citeauthoryear{Xu, Cuesta, Padmanabhan, Eisenstein  \&
  McBride}{Xu et~al.}{2013}]{Xu_2013}
Xu X.,  Cuesta A.~J.,  Padmanabhan N.,  Eisenstein D.~J.,   McBride C.~K.,
  2013, \mn@doi [Monthly Notices of the Royal Astronomical Society]
  {10.1093/mnras/stt379}, 431, 2834Ð2860

\bibitem[\protect\citeauthoryear{Yoo, Fitzpatrick  \& Zaldarriaga}{Yoo
  et~al.}{2009}]{Yoo:2009au}
Yoo J.,  Fitzpatrick A.~L.,   Zaldarriaga M.,  2009, \mn@doi [Phys. Rev.]
  {10.1103/PhysRevD.80.083514}, D80, 083514

\bibitem[\protect\citeauthoryear{Zhang, Liguori, Bean  \& Dodelson}{Zhang
  et~al.}{2007}]{Zhang:2007nk}
Zhang P.,  Liguori M.,  Bean R.,   Dodelson S.,  2007, \mn@doi [Phys. Rev.
  Lett.] {10.1103/PhysRevLett.99.141302}, 99, 141302

\makeatother
\end{thebibliography}

\end{document}